\documentclass[conference, comsoc]{IEEEtran}
\IEEEoverridecommandlockouts

\usepackage{cite}
\usepackage[cmex10]{amsmath}
\usepackage{amsfonts,amsthm,amssymb}
\usepackage{graphicx}
\usepackage{epstopdf}
\usepackage[font=scriptsize,caption=false,labelsep=space]{subfig}
\usepackage{mathtools}
\usepackage{nicefrac}
\usepackage{bm}
\usepackage{txfonts}

\usepackage{balance}

\DeclareRobustCommand*{\IEEEauthorrefmark}[1]{%
\raisebox{0pt}[0pt][0pt]{\textsuperscript{\footnotesize\ensuremath{#1}}}}

\usepackage{upquote}
\usepackage[none]{hyphenat}

\begin{document}
\setlength{\abovedisplayskip}{3pt}
\setlength{\belowdisplayskip}{3pt}

\setlength{\tabcolsep}{5pt}

\newdimen\origiwspc%
\newdimen\origiwstr%
\origiwspc=\fontdimen2\font
\origiwstr=\fontdimen3\font

\title{Beam Alignment and Tracking for Autonomous Vehicular Communication using IEEE 802.11ad-based Radar\vspace{-12pt}}

\author{\IEEEauthorblockN{\fontdimen2\font=0.35ex{\IEEEauthorrefmark{1}Guillem Reus Muns,
\IEEEauthorrefmark{2}Kumar Vijay Mishra, 
\IEEEauthorrefmark{3}Carlos Bocanegra Guerra,
\IEEEauthorrefmark{4}Yonina C. Eldar and
\IEEEauthorrefmark{5}Kaushik R. Chowdhury}\fontdimen2\font=\origiwspc} 
\IEEEauthorblockA{\IEEEauthorrefmark{1,3,5}Electrical and Computer Engineering Department, Northeastern University, Boston, MA, USA\\\IEEEauthorrefmark{2,4}Andrew and Erna Viterbi Faculty of Electrical Engineering, Technion - Israel Institute of Technology, Haifa, Israel\\
Email:
{\{\IEEEauthorrefmark{1}greusmuns@coe,\IEEEauthorrefmark{3}bocanegrac@coe, \IEEEauthorrefmark{5}krc@ece\}.neu.edu}, {\{\IEEEauthorrefmark{2}mishra,\IEEEauthorrefmark{4}yonina\}@ee.technion.ac.il}}}

\maketitle

\begin{abstract}

Mobility scenarios involving short contact times pose a challenge for high bandwidth data transfer between autonomous vehicles and roadside base stations (BS). Millimeter wave bands are a viable solution as they offer enormous bandwidth in the 60GHz band with several Gbps data transfer rates. However, beamforming is used as a default mode in this band, which requires accurate and continuous alignment under relative motion. We propose a method in which an off-the-shelf IEEE 802.11ad WiFi router is configured to serve as the BS as well as a radar exploiting special structure of 802.11ad preamble. We embed the radar functionality within standards-compliant operations that do not modify the core structure of the frames beyond what is defined by the 802.11ad protocol. This not only reduces the beam training time, but also ensures scalability with increasing vehicular traffic because radar allows accurate ranging of up to 0.1m at distances up to 200m. We further analyze the ensuing cost-benefit trade-off between the time allotted to the proposed in-band radar and communication modes. Our results reveal 83\% reduction on the overhead incurred during the beam training achieved for a specific simulated vehicular scenario over the classical 802.11ad operation.
\end{abstract}

\begin{IEEEkeywords}
IEEE 802.11ad, mmWave, automotive radar, beam alignment, CAV
\end{IEEEkeywords}

\IEEEpeerreviewmaketitle
\vspace{-8pt}
\section{Introduction} \label{intro}
Connected and autonomous vehicles are being gradually integrated into consumer-level road transportation given the associated cost/energy savings~\cite{qureshi2013survey}, as well as increase safety and comfort.
In addition, several companies use the same technology to generate digital maps and real-time 3-D visuals through real-time sensing, which can improve offline path planning algorithms~\cite{levinson2011towards}. Each of these examples requires efficient relaying of large volumes of data by the vehicles, ranging in the order of several gigabits-per-second. The key problem that we address in this paper is how to devise a cost-effective communication architecture for vehicle-to-infrastructure (V2I) that offers high data capacity while being resilient to the challenges posed by mobility. 

Millimeter Wave (mmWave) technology in the recently opened $57-64$ GHz band is a promising candidate for our scenario because of the availability of massive unlicensed spectrum. Consumer-exploitation of this band is already underway: The IEEE 802.11ad standard enables throughput of up to 7 Gbps at 60 GHz for short-range wireless communication \cite{ieee2012phy80211ad}. Although the available bandwidth is considerably wider in these bands (for e.g., IEEE 802.11ad defines up to $2$ GHz wide channels), transmissions suffer from high attenuation associated with increasing carrier frequency and channel losses arising from natural phenomena such as atmospheric absorption. In order to overcome these problems, highly directional antennas are used. Here, the radiation beam patterns are configured to ensure the  peak of the transmitted RF energy appears along the intended direction, while lowering the emissions in other locations.
\begin{figure}[t!]
\centering \includegraphics[width=0.475\textwidth]
{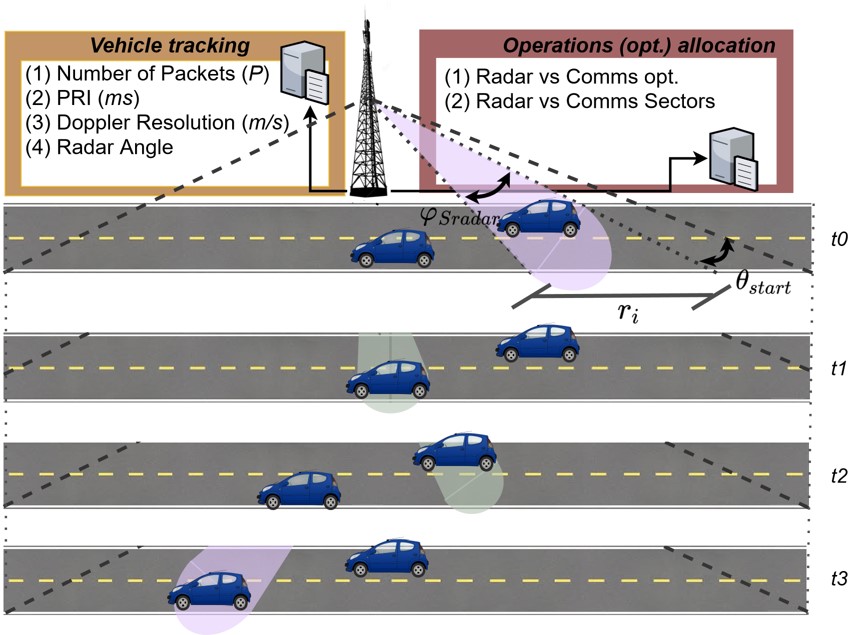}
\caption{\small{V2I scenario where a BS uses radar to localize and serve vehicles moving right to left. Radar operations (purple) allow for determining the location and speed, enabling directive and robust data transfer (green). These operations are scheduled within a standard MAC frame given in Fig.~\ref{fig:MAC_PHY}.}\vspace{-8pt}}
\label{fig:radarBS}
\end{figure}

\noindent$\bullet$ \textbf{Challenges in beamforming for mobile scenarios:} First, narrow beams efficiently focus energy in the chosen direction, but are less robust to movement, i.e. any beam misalignment causes a sharp loss in connectivity. Second, the beam training procedure is time-consuming and lowers efficiency by utilizing network resources that could be used for communication. This worsens in mobile scenarios where the beam training function calls are frequent~\cite{mavromatis2017beam}. Training involves two major steps: through per-sector transmissions, users first locate each other in space during \textit{coarse-grained} training, followed by exploration of the identified sector \textit{fine-grained} training to lower the beamwidth. 
Finally, the use of narrow directional beams makes the connection sensitive to blockage, which is a significant problem in mmWave compared with sub-6 GHz bands. In most mobility scenarios, contact times are expected to be short. Hence, if standards-compliant protocols are used, classical beam training techniques are not feasible. \vspace{4pt}

\begin{figure}[t]
\centering \includegraphics[width=0.495\textwidth]{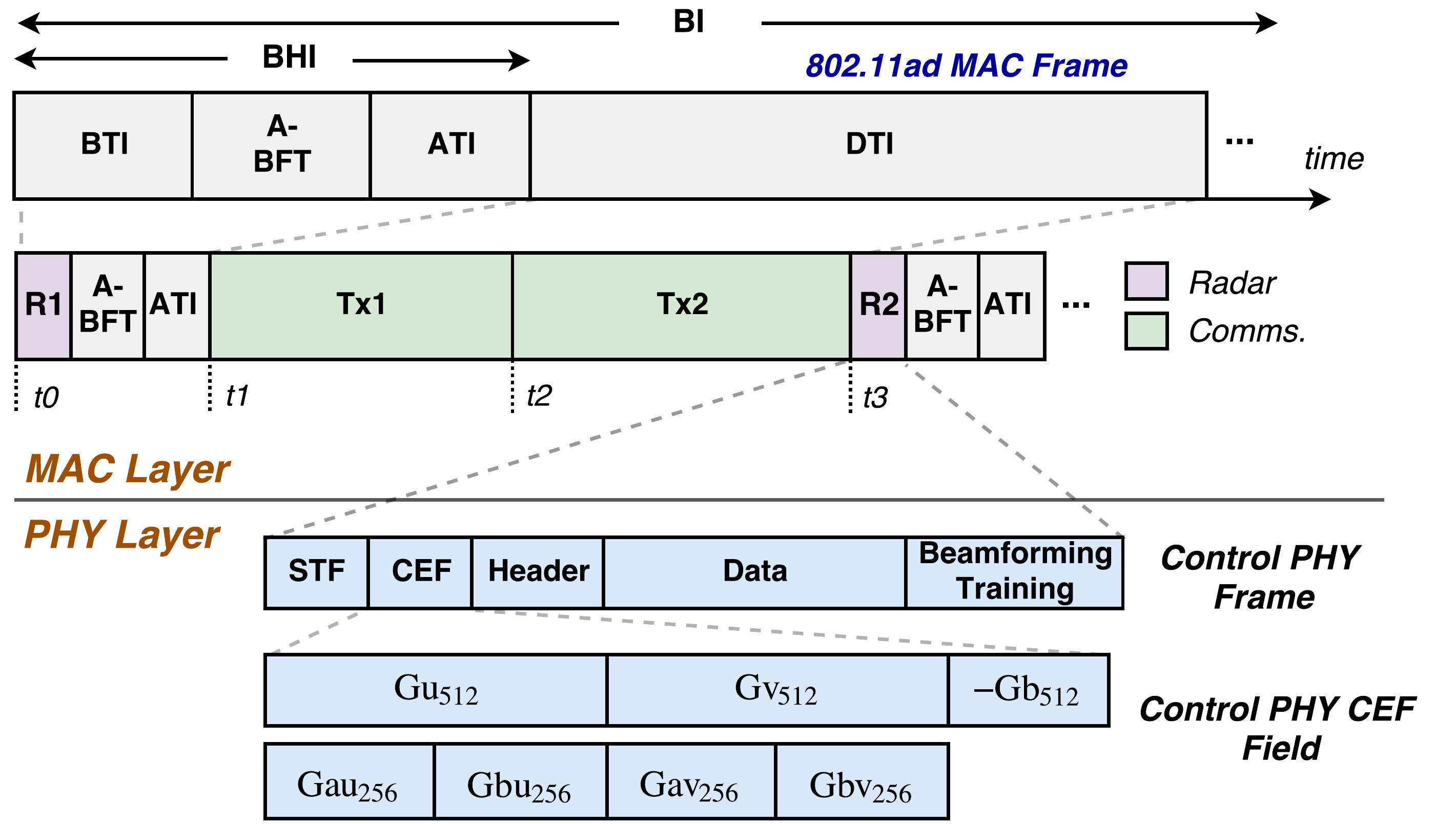}
\caption{\small{The 802.11ad MAC frame configured to accommodate radar operations. R1 and R2 represent the radar periods while Tx1 and Tx2 are for the vehicle data transfer. The details of the physical frame (bottom) IEEE 802.11ad MAC header (top) are described in Sections \ref{sec:radar} and \ref{sec:sys_desc}, respectively.}\vspace{-4pt}}
\label{fig:MAC_PHY}
\end{figure}

\noindent$\bullet$ \textbf{Proposed Approach:} Our approach performs both radar operations and conventional communication using a common transmitter/receiver (Tx/Rx) chain (Fig.~\ref{fig:radarBS}) with 802.11ad physical layer (PHY) frames, as proved earlier in \cite{mishra2017sub,kumari2015investigating}. However, it extends further to the Media Access Control (MAC) layer by integrating these functions within defined MAC protocol headers (see Fig.~\ref{fig:MAC_PHY}), but does not require any protocol modification.
This approach quickly, and in a scalable manner, identifies the locations of moving vehicles, which in turn cuts down the time needed for classical beam training. We address a number of technical issues to achieve this design: (i) incorporating the radar capability within the constraints and features of the protocol standard, (ii) analyzing and optimizing the durations allotted to radar and communications, and (iii) mapping the communications technology advancement to the operational demands of autonomous vehicles.


Our main contributions are:
\begin{enumerate}
\item We design and profile a 802.11ad PHY-based radar in terms of Doppler resolution, which determines the accuracy of the radar, its ability to track the vehicles, and the required time of operation.

\item We connect an in-band radar ranging method with the classical 802.11ad operation, and study the associated cost-benefit trade-offs among the efficiency of channel utilization, throughput, beam dimensions, among others.

\item We propose a 802.11ad MAC configuration to accommodate radar operations alongside regular communications operations, without any change in the channel access operation. Consequently, this reduces misalignment in a high-mobility environment while maintaining low overhead. 

\item We provide spatial characterization of the beamwidth, angular sector and sector overlap as a function of the radar performance, and analyze the resulting throughput when in-band radar and communication operations are performed using the same hardware.

\end{enumerate}


This paper is organized as follows. Section~\ref{sec:related} reviews the state of the art. Section~\ref{sec:radar} describes the technical details of the proposed 802.11ad-based radar, which influences the parameters used in the later sections. Section~\ref{sec:throughput} contains a detailed analysis of the achievable throughput. In Section~\ref{sec:sys_desc}, we provide an operational description of the vehicular scenario and connect the technical development to the application scenario. Section~\ref{sec:res} gives the performance evaluation and, finally, Section~\ref{sec:concl} concludes the paper. 
\vspace{-4pt}
\section{Related Work}
\label{sec:related}
Many recent research efforts have been devoted to reducing the beam training overhead in the mmWave band. In \cite{tsang2011coding}, the authors reduce the time for beam training using in-packet BF. In \cite{weixia2014new}, a codebook design scheme enhances the system performance. The 60 GHz and legacy 2.4/5 GHz bands are combined in \cite{nitsche2015steering} to accelerate the beam training.

While the above mentioned works reduce the overhead of beam training, they are not designed for mobility situations. Recent work involving mobility leverages out-of-band position reports to aid in beam training \cite{mavromatis2017beam,va2016beam}. Similar developments have been published for high speed trains~\cite{va2015beam}. However, this work relies on position reports supplied through other bands to facilitate the directional communications. On the other hand, we use the same (IEEE 802.11ad) band to obtain such reports. 

The IEEE 802.11ad physical layer transmits control (CPHY), single carrier (SC) and orthogonal frequency-division multiplexing (OFDM) modulation frames at chip rates of 1.76 GHz and 2.64 GHz, respectively. Each IEEE 802.11ad CPHY and SCPHY frame consist of a short training field (STF), a channel estimation field (CEF), header, data and a beamforming training field (see Fig.~\ref{fig:MAC_PHY}). The CPHY and SCPHY frames have the same chip rate, carrier frequency and CEF structure. However, the CPHY STF is longer than that of the SCPHY. The STF and CEF together form the corresponding (CPHY or SCPHY) preambles. Within the CEF lie two 512-point sequences $Gu_{512}[n]$ and $Gv_{512}[n]$ which encapsulate Golay pairs. These paired sequences have the property of perfect \textit{aperiodic} autocorrelation (zero sidelobes) making them useful for communication channel estimation \cite{mishra2017sub} and radar remote sensing \cite{kumari2015investigating}. In this paper, our radar employs 802.11ad CPHY/SCPHY frames to estimate the range and Doppler velocity of the vehicles.

Among prior studies, the closest to our work is \cite{gonzalez2016radar} which employs radar for the beam-alignment but it does not specify any particular protocol for communication. In addition, it excludes high-mobility vehicles and the radar does not co-exist within the same band. In \cite{kumari2015investigating}, an in-band radar that uses 802.11ad signaling is presented for vehicle-to-vehicle (V2V) applications. The model assumes a single target and estimates its range using SCPHY STF and symbol boundary detection (SBD), while a coarse estimation of Doppler velocity is obtained using the pulse-pair method \cite{george2010implementation,mishra2012signal}. Similar single-target analyses with CPHY STF are described in \cite{grossi2017opportunistic}.

Very accurate estimates of radar range and Doppler velocity are obtained in \cite{nguyen2017delay} using a $5.89$ GHz dedicated short-range communication (DSRC)-based V2V communication link. However, the signaling is based on the 802.11p OFDM protocol where, for the radar operation, the subcarrier spacing should be more than the maximum Doppler shift, and the cyclic prefix duration needs to be greater than the longest round-trip delay. It is difficult to maintain these requirements in a relatively narrow $10$ MHz DSRC band. In \cite{cohen2016spectrum}, a general-purpose spectrum sharing radar is presented using the recently proposed concept of cognitive radar that exploits sub-Nyquist processing \cite{bar2014sub}. The automotive radar application modifications are suggested in \cite{mishra2017auto} but these works employ IEEE 802.22 cognitive radio for communications.

In contrast to these works, we use 802.11ad-based pulse-Doppler radar for beam alignment and radar network synchronization without any restriction on  mobility. Unlike \cite{kumari2015investigating,grossi2017opportunistic}, our model and processing are also applicable for both CPHY and SCPHY frames.
\vspace{-6pt}
\section{802.11ad V2I Automotive radar} 
\label{sec:radar}
Our range and Doppler estimation methods closely follow the CEF-based communication channel estimation described first in \cite{mishra2017sub,kumari2015investigating} 
and, hence, we only summarize them here.
\vspace{-6pt}
\subsection{Automotive radar signal processing}
\label{subsec:sigpro}
A \textit{Golay complementary pair} consists of two sequences $Ga_N$ and $Gb_N$ both of the same length $N$ with entries $\pm1$. Consequently, their \textit{aperiodic} autocorrelation functions have sidelobes equal in magnitude but opposite in sign. Therefore, the sum of the two autocorrelations has a peak of $2N$ and a sidelobe level of zero:\par\noindent\small
\begin{align}
\label{eq:golaytimeprop1}
Ga_N[n]*Ga_N[-n] + Gb_N[n]*Gb_N[-n] = 2N\delta[n],
\end{align}\normalsize
where $*$ denotes linear convolution. The 802.11ad CEF transmit signal is a concatenated sequence:\par\noindent\small
\begin{align}
\label{eq:txseq}
s_{T}[n] &= Gu_{512}[n] + Gv_{512}[n-512], \phantom{1}n=0, 1, \cdots, 1023.
\end{align}\normalsize
Here, the sequences $Gu_{512}[n]$ and $Gv_{512}[n]$ are defined for $0 \le n \le 511$; for other values of $n$, they are set to zero. Each of them contains a Golay complementary pair of length $256$, $\{Gau_{256}, Gbu_{256}\}$ and $\{Gav_{256}, Gbv_{256}\}$, respectively. The discrete-time sequence $s_T[n]$ is passed through a digital-to-analog-converter (DAC) the output of which can be represented as a weighted sum of Dirac impulses $s_T(t) = \sum\limits_{m}s_T[m]\delta(t-mT_c)$.

The 802.11ad protocol specifies a spectral mask for the transmit signal to limit inter-symbol interference (ISI) \cite[section 21.3.2]{ieee2012phy80211ad}. We assume that $h_T(t)$ includes a low-pass baseband filter with an equivalent amplitude characteristic of the spectral mask. A common shaping filter has a frequency response $H_T(f)$ of the root raised cosine (RRC) filter. The receiver employs another RRC filter $H_R(t)$ such that the net response is equal to a raised cosine (RC) filter, $H(f) = H_T(f)H_R(f)$. The RC filter is a Nyquist filter with the following property to avoid ISI:\par\noindent\small
\begin{align}
\label{eq:rc1}
h[n]=h(nT_c)=\begin{dcases}
1,\;n=0\\
0,\;n\neq 0\\
\end{dcases}.
\end{align}\normalsize
We can formulate this as:\par\noindent\small
\begin{align}
\label{eq:rc2}
h(t)\sum\limits_{k=-\infty}^{+\infty}\delta(t-kT_c)=\delta(t).
\end{align}\normalsize
This property only holds for the RC, and not the RRC filter. The baseband signal is then upconverted for transmission: $x(t) = x_T(t)e^{j2\pi f_ct}$, where $f_c$ denotes the carrier frequency.

Suppose the radar transmits $P$ packets at the pulse repetition interval (PRI) $T_p$ towards $L$ direct-path nonfluctuating targets of complex reflectivity $\alpha_l$ located at range $d_l = c\tau_l/2$ and Doppler velocity $\nu_l = 2\pi f_{D_l}$, where $c = 3\times10^8$ m/s is the speed of light, $\tau_l$ is the time delay, and $f_{D_l}$ is the associated Doppler frequency. Ignoring the multi-path components (MPC), the reflected received signal at the baseband, i.e., after down-conversion, is given by\par\noindent\small
\begin{align}
x_R(t) &= \sum\limits_{p=0}^{P-1}\sum\limits_{l=1}^L \alpha_l x_T(t-\tau_l-pT_p)e^{-j2\pi f_{D_l} t} + z(t)\nonumber\\
&\approx \sum\limits_{p=0}^{P-1}\sum\limits_{l=1}^L \alpha_l x_T(t-\tau_l-pT_p)e^{-j2\pi f_{D_l}pT_p} + z(t),
\end{align}\normalsize
where $z(t)$ is additive circular-symmetric white Gaussian noise. The last approximation follows from the fact that $f_{D_l} \ll 1/T_p$ so that the phase rotation within one coherent processing interval (CPI) (``slow time'') can be approximated as a constant. Here, we have assumed that the coefficient $\alpha_l$ also characterizes antenna directivity, processing gains and attenuations including path loss. 

The received signal is sampled at $F_c = 1/T_c$ to obtain\par\noindent\small
\begin{align}
x_R[n] &= x_R(nT_c) = \sum\limits_{p=0}^{P-1}\sum\limits_{l=1}^L \alpha_l x_T(nT_c-\tau_l-pT_p)e^{-j2\pi f_{D_l}pT_p} + z(nT_c)\nonumber\\
&= \sum\limits_{p=0}^{P-1}\sum\limits_{l=1}^L \alpha_l s_T(nT_c-\tau_l-pT_p)e^{-j2\pi f_{D_l}pT_p} + z[n],
\end{align}\normalsize
where we used the Nyquist filter properties (\ref{eq:rc1})-(\ref{eq:rc2}) to obtain the last equality. The sampled signal is passed through matched filters of each Golay sequence e.g. for the first pair, we have\par\noindent\small
\begin{align}
\hat{h}_{1au}[n] &= x_R[n]*Gau_{256}[-n]\nonumber\\
\hat{h}_{1bu}[n] &= x_R[n]*Gbu_{256}[-n].
\end{align}\normalsize
These outputs are delayed: $\hat{h}_{1a}[n] = \hat{h}_{1au}[n]$, $\hat{h}_{1b}[n] = \hat{h}_{1bu}[n+256]$, and summed up to yield the channel estimate\par\noindent\small
\begin{align}
\hat{h}_{1}[n] &= \frac{1}{512}(\hat{h}_{1a}[n] + \hat{h}_{1b}[n])\nonumber\\
			   &\approx \frac{1}{512}\sum\limits_{p=0}^{P-1}\sum\limits_{l=1}^L \alpha_l \delta(nT_c-\tau_l-pT_p)e^{-j2\pi f_{D_l}pT_p}\nonumber\\
               &+ z[n]*(Gau_{256}[-n] + Gbu_{256}[-n]),
\end{align}\normalsize
where we assumed that the Doppler shifts are nearly identical for the two Golay sequences $Gau_{256}$ and $Gbu_{256}$ enabling us to use the Golay pair property (\ref{eq:golaytimeprop1}). Another estimate $\hat{h}_2[n]$ of the radar channel is available via the second Golay pair within $Gv_{512}$. An average of the two estimates yields a final approximation of the radar channel. 

We discretize the range-time space in, say, $N_R$ bins of resolution $cT_c/2$. We then create a delay-Doppler map by taking a $P$-point Discrete Fourier Transform (DFT) of the radar channel estimates corresponding to each delay bin. Then, the delay and Doppler frequencies of the targets are given by the location of the first $L$ peaks on this 2D delay-Doppler map using e.g. a constant false alarm rate (CFAR) detector. In a pulse Doppler radar, the above-mentioned processing estimates only the radial component of the vehicle's velocity. If needed, a radar can estimate the tangential component by measuring target's motion on a clutter map. 
\vspace{-6pt}
\subsection{V2I radar design}
\label{subsec:v2iradar}
At the carrier frequency $f_c$ of $60$ GHz, the wavelength is $\lambda_c = c/f_c = 0.005$ m. The STF of CPHY and SCPHY frames contain $6400$ and $2176$ symbols, respectively. The CEF of both frames has $1152$ symbols each. For the symbol time of $T_c = 1/1.76e9 = 0.568$ ns, we obtain a minimum PRI of $T_{pr} = T_c*(6400+1152) = 4.29$ $\mu s$ for CPHY while the same is $1.89$ $\mu s$ for the SCPHY. A typical long range V2I radar (LRR) operates with a maximum unambiguous range of $R_u = 200$ m and range resolution of $0.1$ m. The symbol time of 802.11ad yields a good range resolution of $\Delta R = cT_c/2 = 8.52$ cm.

\begin{figure}[t!]
\centering
	\subfloat[]{%
  	\includegraphics[width=0.46\linewidth]{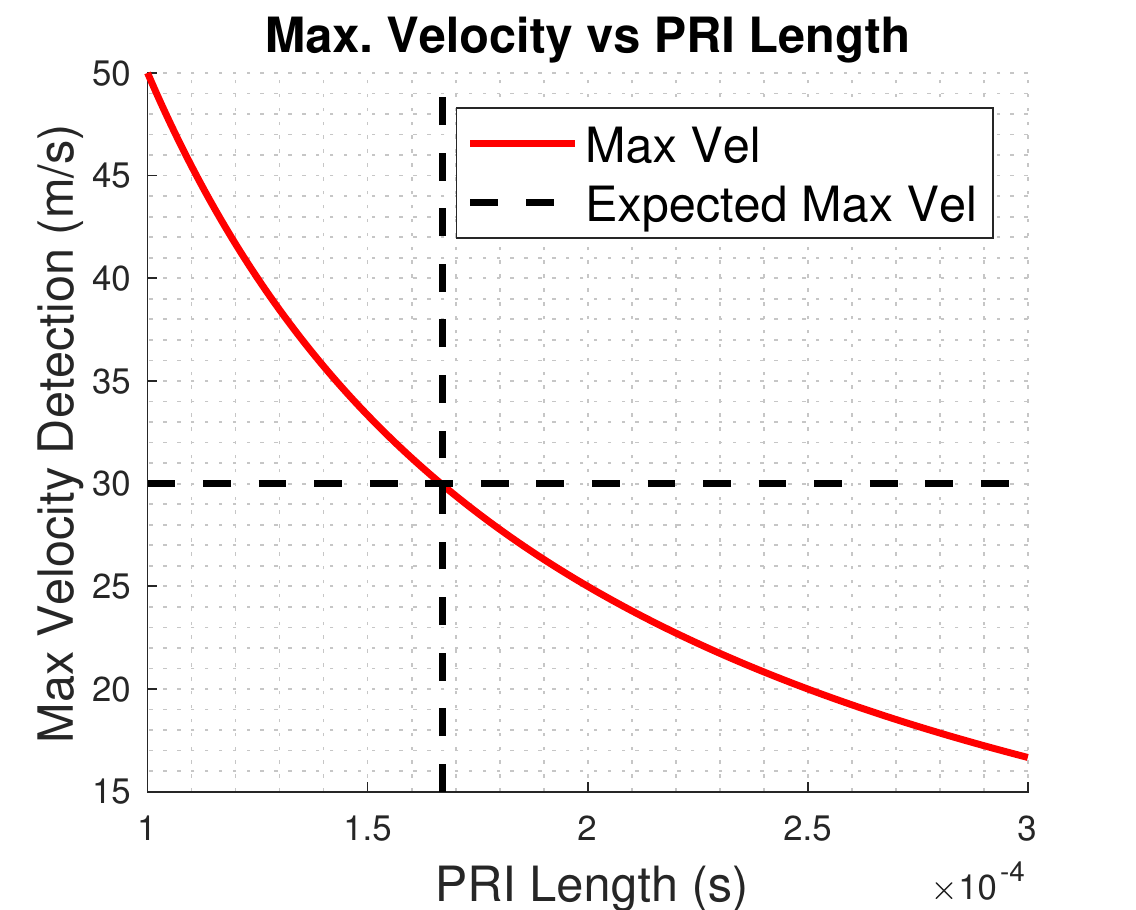}%
  	\label{subfig:MaxSpeed}%
	}\qquad
	\subfloat[]{%
  	\includegraphics[width=0.46\linewidth]{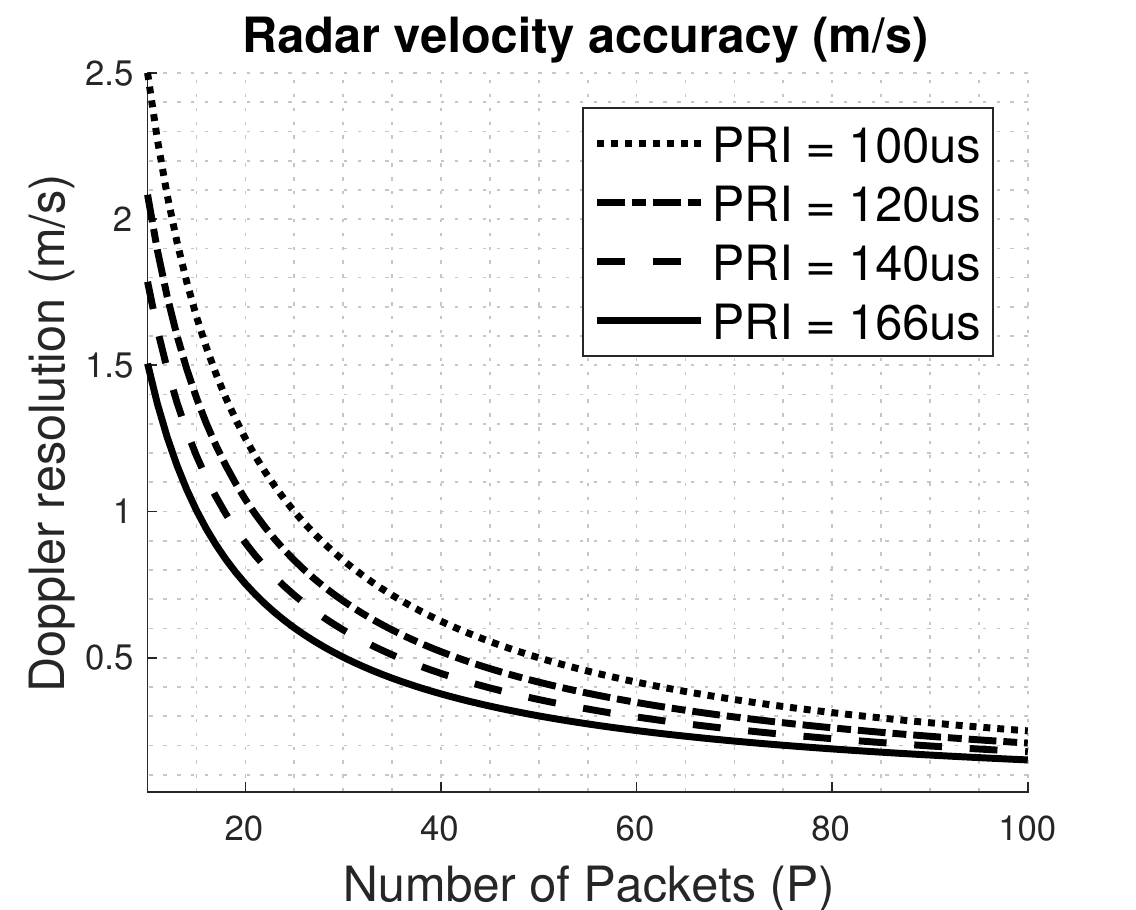}%
  	\label{subfig:tDopplerResol}%
	}
\caption{\small{(a) Maximum detectable velocity $\nu_u$ as a function of the PRI length. For highways, where $\nu_u = 30$ m/s, PRI shall be no greater than 0.166 ms. (b) The radar accuracy as a function of \textit{P} for PRI no greater than the highway limit.}\vspace{-4pt}}
\label{fig:dopres}
\end{figure}

If $P$ packets are transmitted, then the CPI duration is $T_{\text{int}} = PT_{pr}$. For pulse repetition at a uniform rate, the maximum unambiguous Doppler velocity of the radar is given by $\nu_u = \lambda_c/T_{pr}$ and the Doppler resolution obtained through Fourier processing is $\Delta\nu = \lambda_c/(2T_{\text{int}})$. Fig.~\ref{fig:dopres} shows the dependence of the Doppler resolution on the number of packets and PRI. 
The total time the radar requires to scan a certain radar sector can be expressed as:\par\noindent\small
\begin{align}
\label{eq:tradar}
t_{\text{radar}} \approx T_{\text{int}}\lfloor \varphi_{\text{Sradar}}/\theta_{\text{radar}} \rfloor,
\end{align}\normalsize
where $\lfloor\cdot\rfloor$ denotes the floor function, $\theta_{\text{radar}}$ is the scan rate per unit CPI, fixed to $0.5^{\circ}$/CPI, and $\varphi_{\text{Sradar}}$ is the angular sector size.
\vspace{-4pt}
\section{802.11ad PHY/MAC Protocol Configuration}
\label{sec:throughput}
\begin{figure}[t]
\centering \includegraphics[width=0.4\textwidth]{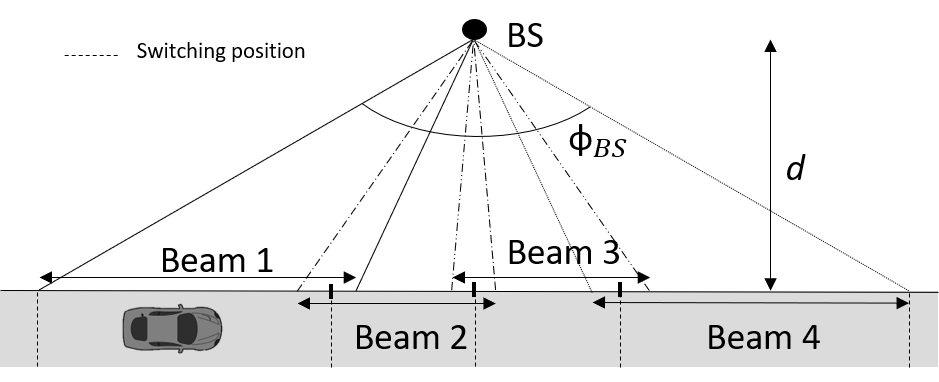}
\caption{\small{Example of beam design with four beam.}\vspace{-4pt}}
\label{fig:scenario}
\end{figure}

Consider a BS located at distance $d$ from an ideal straight road, serving $V$ number of non-stationary vehicles that pass along a fixed and uniform route (Fig.~\ref{fig:scenario}). Each vehicle $c_l$ is characterized by its true location and velocity at time $t$, $y_l(t)$ and $\nu_l(t)$, respectively. The BS obtains the radar estimates $\hat{d}_l(t)$  and $\hat{\nu}_l(t)$) of the location and velocity, respectively. For simplicity, $\hat{\nu}_l(t)$ will be assumed to be constant. As we describe below, the 802.11ad throughput is affected by the location of the vehicles with respect to the BS. 
\subsection{Link Budget Model} \label{subsec:link}
Given the communication scenario (Fig.~\ref{fig:radarBS}) and transmit power $P_{\text{tx}}$, we assume line-of-sight (LOS), allowing for the following simplified expression for the received power $P_{\text{rx}}$ \cite{mavromatis2017beam} (all quantities in dB): 
\par\noindent\small
\begin{align}
\label{eq:TH_RxPower}
P_{\text{rx}} = P_{\text{tx}} + G_{\text{tx}} - PL + G_{\text{rx}},
\end{align}\normalsize
where $PL$ is the path loss and $G_{\text{tx}}$ and $G_{\text{rx}}$ are the antenna gains in transmission and reception, respectively. Here, the path loss is given by\par\noindent\small
\begin{align}
\label{eq_TH_PL}
PL = 10n\log_{10}d_v + SF + C_{\text{att}} + A_{\text{att}} + R_{\text{att}},
\end{align}\normalsize
where $n$ is the path loss exponent, $d_v$ is the distance from the Tx in the BS to the vehicle, and $SF$ corresponds to the random shadowing effect following a lognormal distribution $SF\sim\mathcal{N}(0,\sigma^2)$ with $\sigma = 5.8$. The $C_{\text{att}}$, $A_{\text{att}}$ and $R_{\text{att}}$ are the LOS link, atmospheric and rain attenuations, respectively. The linear scale antenna gains are given by\par\noindent\small
\begin{align} 
\label{eq_TH_GainIdeal}
G_{\text{tx}/\text{rx}} = \frac{4\cdot 180^2 }{\theta_{\text{el}} \theta_{\text{az}} \pi},
\end{align}\normalsize
where $\theta_{\text{el}}$ and $\theta_{\text{az}}$ are the half-power beamwidths (radians) in elevation and azimuth for the corresponding antennas. To be consistent with the previous work in \cite{mavromatis2017beam} and \cite{va2016beam}, we assumed an ideal beamforming in (\ref{eq_TH_GainIdeal}), characterized by uniformly distributed gain within the beam and no side lobes (ideal antenna efficiency).


We model the impairments at the receiver by accounting for the phase noise $P_{\text{noise}}$ as follows:
\par\noindent\small
\begin{align}
\label{eq:TH_PN}
P_{\text{noise}} = N_{\text{floor}} + 10\log_{10}B + \text{NF},
\end{align}\normalsize
where $N_{\text{floor}}$ is the noise floor, $B$ is the system bandwidth and $NF$ the noise figure of the receiver chain.
As a result, the SNR is a function of the Tx-Rx distance $d_v$ and beamwidth:\par\noindent\small
\begin{align}
\label{eq:TH_SNR}
\text{SNR}(\theta_{\text{az}}, d_v) = \frac{P_{\text{rx}}(\theta_{\text{az}}, d_v)}{P_{\text{noise}}}.
\end{align}\normalsize
Here, $\theta_{\text{el}}$ is fixed to a value that encompasses the width of the road. Hence, it depends on the distance $d$. Both $\theta_{\text{el}}$ and $\theta_{\text{az}}$ are assumed identical in transmission and reception.




\vspace{-6pt}
\subsection{Average Rate}
\label{subsec:per}
For each Modulation and Coding Scheme (MCS), the 802.11ad standard requires a minimum Packet Error Rate (PER) which, in turn, is inherently impacted by the Signal-to-Noise Ratio (SNR). As in (\ref{eq:TH_SNR}), the $\text{SNR}$ itself depends on the scenario geometry $d_v$ and $\theta_{\text{az}}$. We can express this dependency as a function $\text{SNR}\left(\theta_{\text{az}}, d_v \right)$. Further, the MCS-0 (most robust) demands a PER lower than a threshold ($\gamma$) of 5\% for a PHY service data unit (PSDU) length of 256 octets. For any other MCS, the PER shall be less than $\gamma = 1\%$ for a PSDU length of 4096 octets \cite{ieee2012phy80211ad}. Thus, the achievable throughput $R_{802.11ad}$ is characterized by the tuple $\{\text{SNR,PER,MCS}\}$. To obtain the maximum data rate $R$, our approach always selects the MCS corresponding to the highest data rate among the ones that meet the PER requirements:
\par\noindent\small
\begin{align}
\label{eq:TH_RateRealistic}
     R( \theta_{\text{az}}, d_v) = \max_{%
       \substack{%
         \text{MCS} \\
         \text{PER}  \left(\text{SNR}\left(\theta_{\text{az}}, d_v \right)\right)  \leq  \gamma
       }
     }
     R_{802.11ad}(\text{MCS}).
\end{align}\normalsize

The average data rate $\overline{R_l}$, or simply $\overline{R}$, between, say, a $l$th vehicle and the BS is computed as the integral of the achievable rate over the contact time interval [$t_1$, $t_2$]:\par\noindent\small
\begin{align}
\label{eq:TH_Rate1}
\overline{R} = \frac{1}{t_2-t_1} \int_{t_1}^{t_2} R(t) dt.
\end{align}\normalsize
For ease of notation, we rewrite [$t_1$, $t_2$] as [$t_{\text{init}}$, $t_{\text{init}}+t_c$]:
\par\noindent\small
\begin{align}
\label{eq:TH_Rate2}
\overline{R} = \frac{1}{t_c} \int_{t_{\text{init}}}^{t_{\text{init}}+t_c } R(t) dt,
\end{align}\normalsize
where $t_{\text{init}}$ corresponds to the instant when the $l$th vehicle enters the area covered by the BS and $t_c$ is the contact duration: 
\par\noindent\small
\begin{align}
\label{eq:TH_Rate3}
t_c = \frac{2d \tan{\left(\frac{\phi_{\text{BS} }}{2}\right)}}{\nu_l},
\end{align}\normalsize
where $\phi_{\text{BS}}$ is the BS coverage angle (see Fig.~\ref{fig:scenario}).

\begin{figure}[t]
	\centering
	\subfloat[]{%
  	\includegraphics[width=0.46\linewidth]{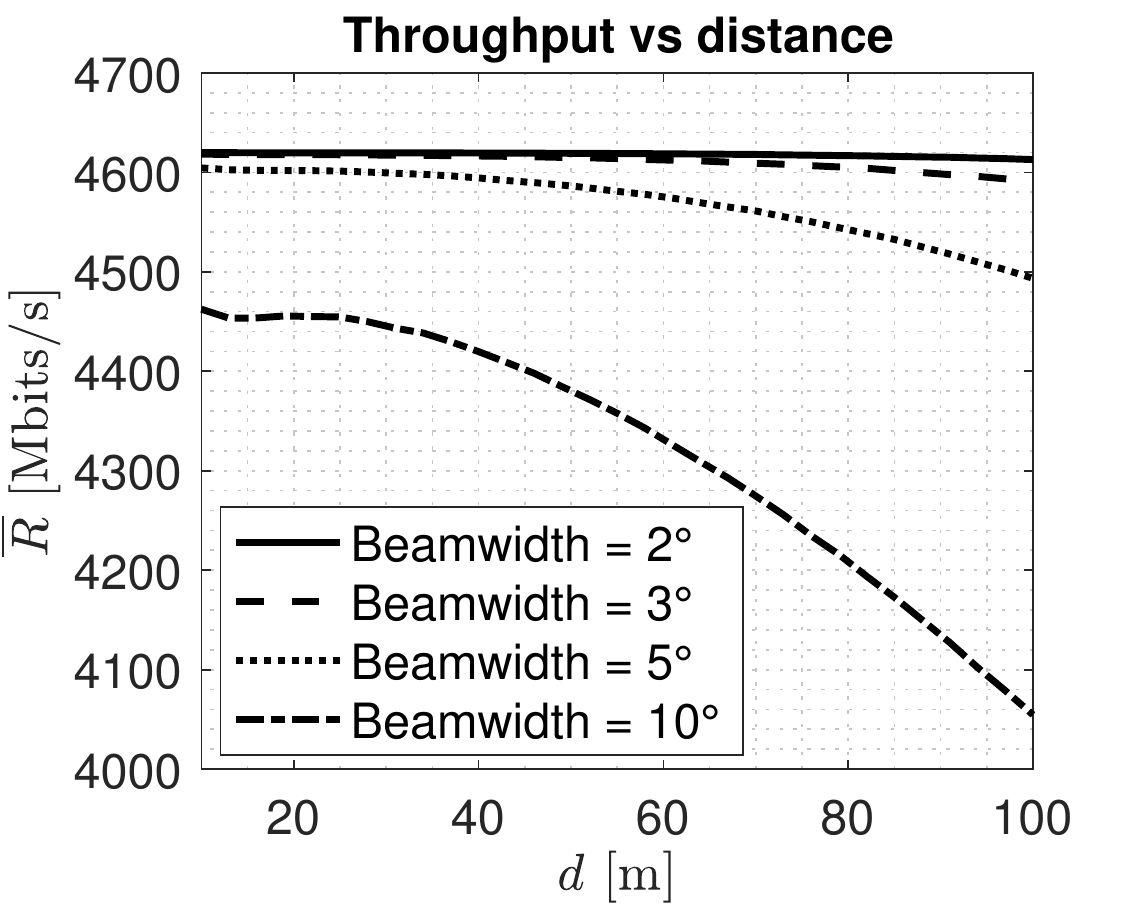}%
  	\label{subfig:th_vs_bw_SC}%
	}\qquad
	\subfloat[]{%
  	\includegraphics[width=0.46\linewidth]{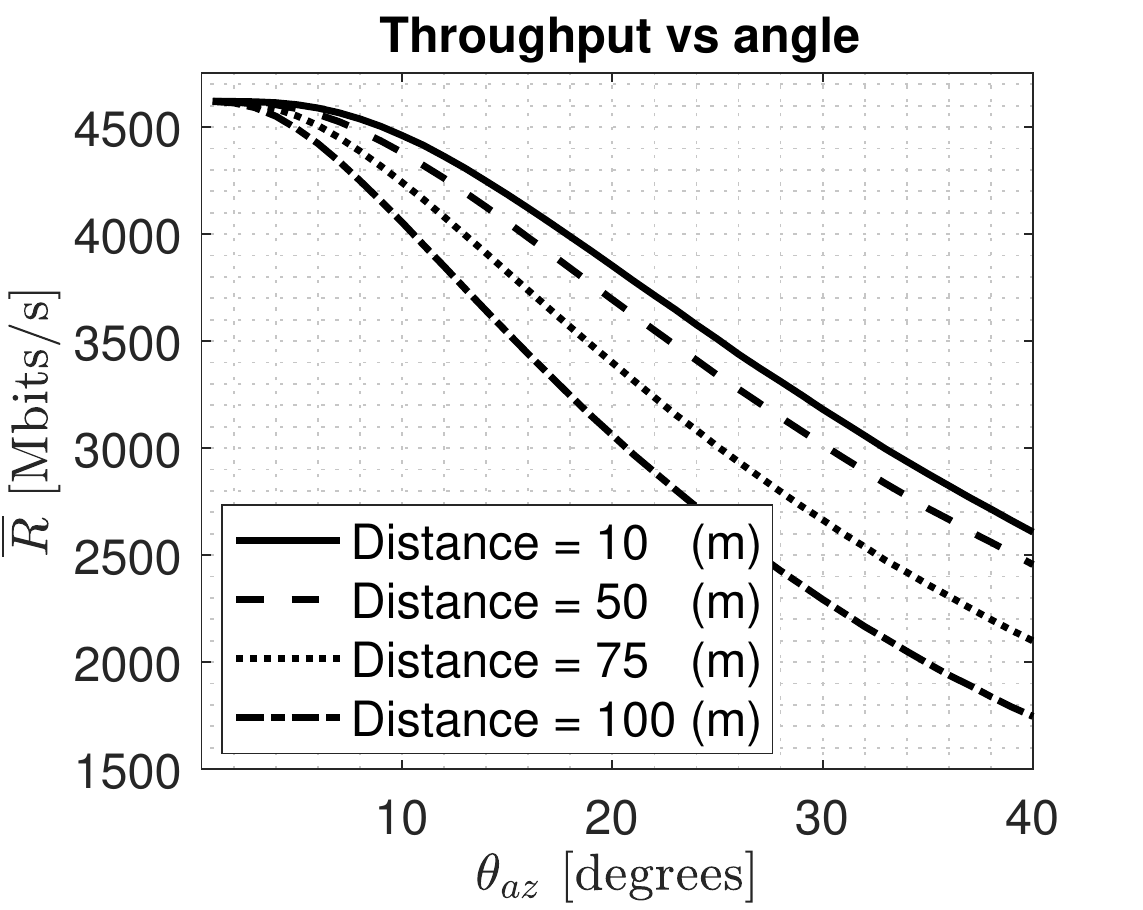}%
  	\label{subfig:th_vs_d_SC}%
	}
\caption{\small{Average throughput analysis with respect to the (a) BS-Road distance and (b) beamwidth.}\vspace{-4pt}}
\label{fig:thrgs}
\end{figure}

From Fig. \ref{fig:thrgs}, where $\overline{R}$ is plotted as a function of $d$ and $\theta_{\text{az}}$, we conclude following: First, a narrower beamwidth helps mitigate the high path loss attenuation, achieve a higher SNR and, in turn, increase the achievable throughput. Second, the achievable throughput is impacted more by the beamwidth $\theta_{\text{az}}$ than the distance $d$. The latter finding serves as a design constraint in our system, i.e., \textit{narrow beamwidth is preferred over short distances}.

\vspace{-4pt}
\section{802.11ad-based Joint radar and Communication}  
\label{sec:sys_desc}
In this section, we discuss the intervals of calling the radar function, and the time alloted to complete this function per call. Beginning with the fields used in the 802.11ad standard, we identify the  elements in the header that should be configured to incorporate the radar operation. 
\vspace{-6pt}
\subsection{Beam alignment model}
\label{subsec:misalign}
Consider the scenario in Fig.~\ref{fig:scenario}. The spatial orientation of the beams follow the beam-switching pattern \cite{va2015beam}. Only one sector is activated at a time, and the BS either performs the radar or communication operations~\cite{va2016beam}. However, the respective sectors are defined differently: For communication, the adjacent sectors have some overlap that is specified by an overlap ratio parameter. The radar sectors do not overlap and their angular aperture is controlled by $\varphi_{Sradar}$. As stated in Section~\ref{subsec:link}, the beams are assumed to be ideal, i.e., if at any instant a vehicle is not within the limits of the active sector, we consider the beams to be in complete misalignment.


\vspace{-6pt}
\subsection{Using 802.11ad Headers}
\label{subsec:MACad}

The 802.11ad protocol structures its frames into Beacon Intervals (BI) (Fig.~\ref{fig:MAC_PHY}) with 30 ms duration for vehicular networks \cite{mavromatis2017beam}. 
The BIs consist of Beacon Header Interval (BHI), where antenna beams are formed and network synchronization information is exchanged, and Data Transfer Interval (DTI), during which the actual communication occurs. On top of this, the protocol also defines a Beam Refinement Protocol (BRP) during the DTI, where a similar procedure is carried out with narrower beams. 
Three subintervals constitute BHI in the following order:
\begin{itemize}
\item \textit{Beacon Transmission Interval (BTI)}. The BS uses beacon frames to announce its presence, scan the whole coverage area using $S_I$ sectors, allow the devices to align their beams and acquire basic network information. This is the initiation phase in a Sector Level Sweep (SLS).
\item \textit{Associated Beamforming Training (A-BFT)}. The devices contend in a random slotted access fashion upon completion of the BTI and the Medium Beamforming Interframe Spacing (MBIFS). The devices first use $S_R$ sectors to reply to the BS, make way to a later feedback message from the BS and send the acknowledgement (ACK) that closes the SLS (Response Phase).
\item \textit{Announcement Transmission Interval (ATI)}. To allocate the devices during the DTI, the BS updates the number of connections and inquires them about their data needs.
\end{itemize}

We explain our PHY-MAC configuration through a scenario requiring vehicles to synchronize with a given BS that has radar capabilities. First, the BS sends $P$ Directional Multi-Gigabit (DMG) Beacon frames (see Section~\ref{subsec:v2iradar}) through $S_I = \varphi_{\text{Sradar}}/\theta_{\text{radar}}$ sectors within a confined area (shown by \texttt{Radar} in Fig.~\ref{fig:radarBS}), allowing the BS to obtain an estimation of the locations and velocities while providing the vehicles with synchronization information. The duration of the \texttt{radar} slot is characterized in (\ref{eq:tradar}) and (\ref{eq:rho1}). At the end of the BTI and after the radar call, the BS detects each car location, and then sends a DMG beacon to each vehicle assigning them to an A-BFT slot (in order to avoid inefficient contention during legacy 802.11ad). The BS antenna is also included in this message, which allows the vehicles to calculate the best transmit sector. Upon completion of the BTI, vehicles start the Response Phase and transmit through their $S_R$ sectors. However, they will already know the BS location, therefore $S_R$ is equal to 1.

In summary, we perform two steps with only one transmission (radar and synchronization) during the BTI. Furthermore, we reduce the A-BFT slots by confining the BTI within the area defined by $\varphi_{Sradar}$ and eliminate the BRP through the high localization accuracy acquired by the radar in the DTI period.

\begin{table}[t]
\caption{Parameter values for numerical experiments}
\label{tab:param}
\begin{tabular}{r l | r l}
\textbf{Parameters} && \textbf{Value} \\
\hline
\hline
Transmission power & $P_{tx}$ & 10 & dBm \\
Path-Loss exponent & $n$ & 2.66 \\
Channel attenuation & $C_{\text{att}}$ & 70 & dB \\
Atmospheric attenuation & $A_{\text{att}}$ & 15 & dB km$^{-1}$ \\
Rain attenuation & $R_{\text{att}}$ & 25 & dB km$^{-1}$ \\
Noise Floor & $N_{\text{floor}}$ & -174 & dBm \\
Bandwidth & $B$ & 2.16 & GHz \\
Carrier frequency & $f_c$ & 60 & GHz \\
Wavelength & $\lambda_c$ & 5 & mm\\
Noise figure & $NF$ & 6 & dB \\
Max. Velocity & $v_{\text{max}}$ & 30 & m/s\\
BS coverage angle & $\phi_{\text{BS}}$ &120& degrees\\
BS-Road distance & d & 100 & m\\
Communications beamwidth & $\theta_{\text{az}}$ & 3 & degrees \\
Outage & & 2 & \% \\
Initiator number of sectors (802.11ad)& $S_I$ & 32&\\
Responder number of sectors (802.11ad)& $S_R$ & 32&\\
Number of A-BFT slots &&4&\\
\vspace{-24pt}
\end{tabular}
\end{table}
\vspace{-4pt}
\subsection{Radar comprehensive configuration - system design}
\label{subsec:radarT}
The radar detects vehicles passing by a pre-defined sector and estimates their position and velocity, $\hat{d}_l(t)$  and $\hat{\nu}_l(t)$, at a given time $t$. In this section, we detail (i) setting up the periodic sweep interval between consecutive sweeps, (ii) selecting sectors for the radar sweep, and (iii) allotting the time per sweep, i.e., decide the number of packets $P$ that should be transmitted to achieve a desirable accuracy.


\noindent $\bullet$ \textbf{Sweep interval:} The minimum elapsed time, $T_{\text{radar},i}$, between two consecutive sweeps across the same $i$th sector, must be set such that every vehicle which passes over that sector is detected. $T_{\text{radar},i}$ (\ref{eq:T}) is restricted by the sector length, $r_i$, and the maximum velocity expected in a vehicle, $v_{\text{max}}$. 

\par\noindent\small
\begin{align}
T_{\text{radar},i} \leq \frac{r_i}{v_{\text{max}}}.
\label{eq:T}
\end{align}\normalsize

\noindent $\bullet$ \textbf{Sector location:} The radar sector length, $r_i$ (\ref{eq:radarArea}), is defined as the road length covered by the radar, which can be expressed as a function of $\varphi_{\text{Sradar}}$, $\phi_{\text{BS}}$ and the radar sector starting angle, $\theta_{\text{start}}$ (see Fig.~\ref{fig:radarBS}).

\par\noindent\small
\begin{align}
\label{eq:radarArea}
r_i = d \left[ \tan\left(\frac{1}{2}\phi_{\text{BS}} - \theta_{\text{start}} \right) - \tan\left(\frac{1}{2}\phi_{\text{BS}} - \left(\theta_{\text{start}} + \varphi_{\text{Sradar}}\right) \right) \right].
\end{align}\normalsize

We simplify the analysis by restricting the number of radar sectors to one. Thus, $\theta_{\text{start}}$ and $\varphi_{\text{Sradar}}$ have to be determined for that single sector. Also, as vehicles must be detected as soon as they enter the coverage area for efficient resource utilization, consider the case where the radar sector is defined by $\theta_{\text{start}} = 0$. Now, the system needs to impose new restrictions towards setting $\varphi_{\text{Sradar}}$: (i) the sector length covered by radar is not smaller than the length of the vehicle, $w_{\text{car}}$ (\ref{eq:lengthcar}), and (ii) the distance that a vehicle may move during a radar sweep cannot be larger than a set value ($k_2$) (\ref{eq:poscar}). The location of the $l$th vehicle at a given time $t$ is denoted by $y_l(t)$. Two  qualitative constraints defined by constants $k_1$ and $k_2$ are included below for flexibility in the analysis.

\par\noindent\small
\begin{align}
r > k_1 \cdot w_{\text{car}},
\label{eq:lengthcar}
\end{align}
\begin{align}
[y_l(t+t_{\text{radar}}) - y_l(t)] < k_2.
\label{eq:poscar}
\end{align}\normalsize

Consider now that the radar is configured to scan the whole coverage area at once. If we want to achieve a Doppler resolution of 1.5 m/s, incurring on a 5\% error on the velocity estimation in the worst case, the system would require around $400$ ms to complete the beam alignment procedure. This is not viable, since a vehicle that was detected at the very beginning would have covered $12$ m upon completion of the radar function (\ref{eq:poscar}). Therefore, the radar location design took a less simplistic approach, employing the radar only in a reduced area.

Consider a $\varphi_{\text{Sradar}}$ of 5$^\circ$, a $\Delta \nu = 0.454$ m/s and a $w_{\text{car}} \approx 5$m. This values would correspond to $r \approx 30$ m ($\approx 6w_{\text{car}}$)  and an in-radar-measure moved distance around 1.75 m ($\approx w_{\text{car}}/3$). This is a numerical example of fair values for $\varphi_{\text{Sradar}}$ and $\Delta \nu = 0.454$ m/s according to the qualitative constraints (\ref{eq:lengthcar}) and (\ref{eq:poscar}).

\noindent $\bullet$ \textbf{Sweep time:} We define $\rho$ as the ratio between the time taken to perform a full sweep of the chosen sector, $t_\text{{radar}}$, and the overall interval between sweeps $T_\text{{radar}}$: 
\par\noindent\small
\begin{align}
\label{eq:rho1}
\rho &= \frac{t_\text{{radar}}}{T_\text{{radar}}},\\
\label{eq:rho2}
\rho(\varphi_{\text{Sradar}}, \theta_{\text{start}}) &= \frac{\varphi_{\text{Sradar}} \cdot \lambda_c \cdot v_\text{{max}}}{\Delta \nu \cdot r(\varphi_{\text{Sradar}}, \theta_\text{{start}}) },
\end{align}\normalsize
where the last equality is obtained by substituting (\ref{eq:tradar}) and (\ref{eq:T}) into (\ref{eq:rho1}).
\begin{figure}[!t]
\centering
\includegraphics[width=0.4\textwidth]{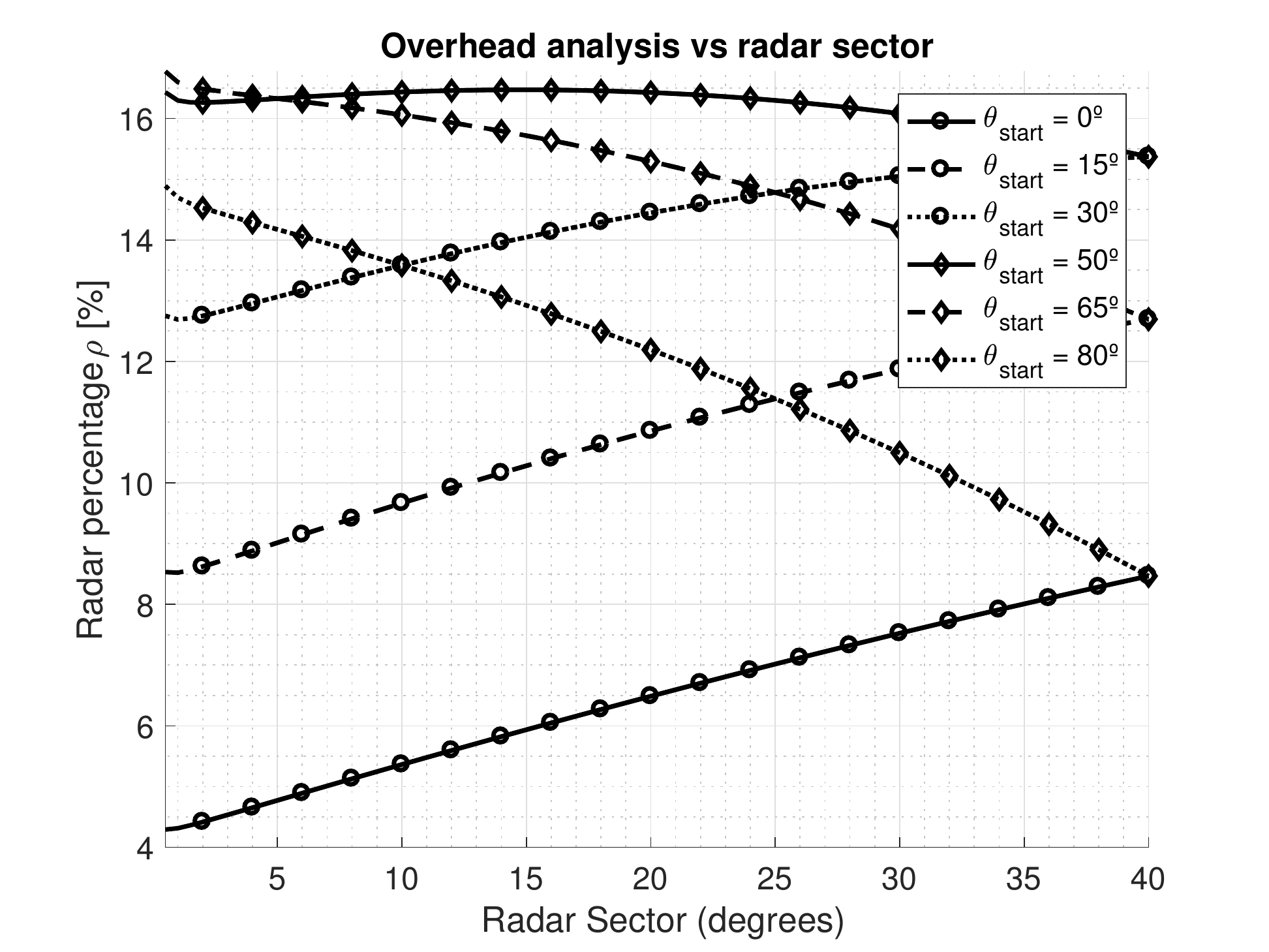}
 \caption{\small{Radar percentage for different starting angles. $\Delta \nu$ = 0.456 m/s is used.}\vspace{-4pt}}
 \label{fig:rho}
\end{figure}
\vspace{-4pt}
\section{Performance Evaluation}
\label{sec:res}
Consider the scenario defined in Section~\ref{subsec:misalign} and represented in Fig.~\ref{fig:scenario}, where vehicles will synchronize with the BS through the procedure defined in Section~\ref{subsec:MACad}. We simulated this operation in MATLAB using the PHY parameters defined in the WLAN Systems Toolbox and Table~\ref{tab:param}. We compared the overhead between the legacy 802.11ad and our modified configuration using 802.11ad-based Radar. 


Fig.~\ref{fig:rho} shows $\rho$ as a function of $\varphi_{\text{Sradar}}$ for different fixed initial locations $\theta_{\text{start}}$. The configuration that provides highest radar efficiency by minimizing $\rho$ is $\theta_{\text{start}}$ = 0. As mentioned in Section~\ref{subsec:radarT}, this aids in the detection of cars as soon as they enter the coverage area.
Therefore, placing the radar sector at the edge of the coverage area is the optimal choice.  

\begin{figure}[t!]
\centering
	\subfloat[]{%
  	\includegraphics[width=0.45\linewidth]{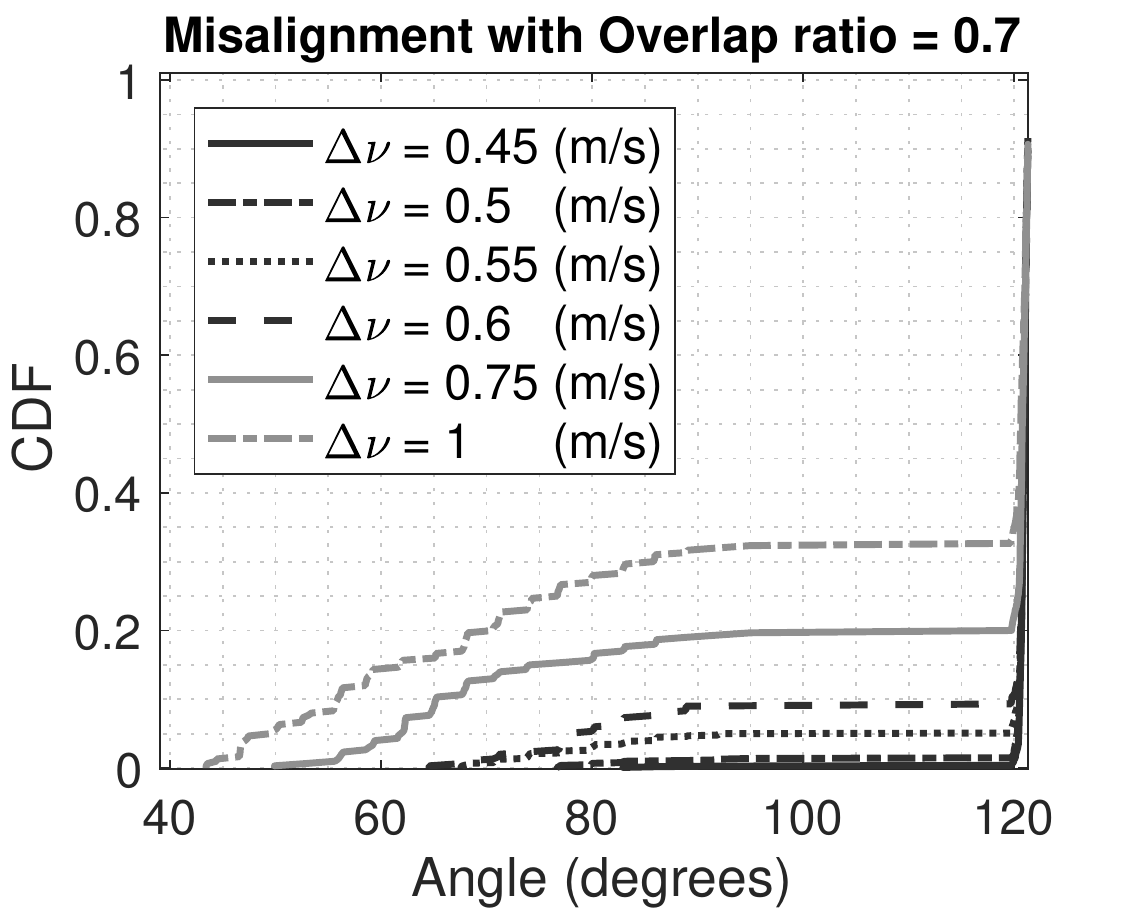}%
  	\label{subfig:cdf_overlap07}%
	}\qquad
	\subfloat[]{%
  	\includegraphics[width=0.45\linewidth]{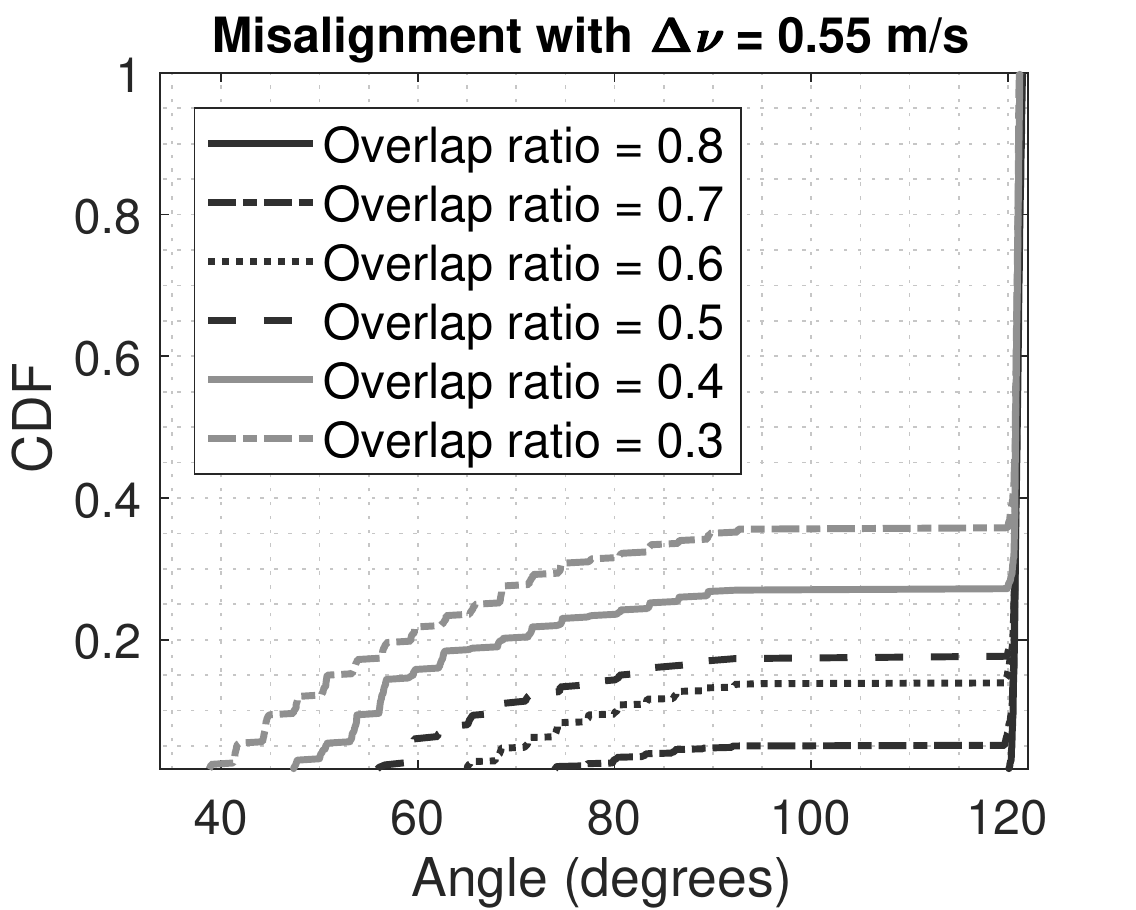}%
  	\label{subfig:cdf_rho_dopp}%
	}
\vspace{1pt}
\caption{\small{Misalignment cumulative probability density function (CDF) for (a) different Doppler resolutions and (b) overlap ratios. The outage tolerance is 2\%.}}
\label{fig:CDFs}
\end{figure}

\begin{figure}[t!]
\centering
	\subfloat[]{%
  	\includegraphics[width=0.45\linewidth]{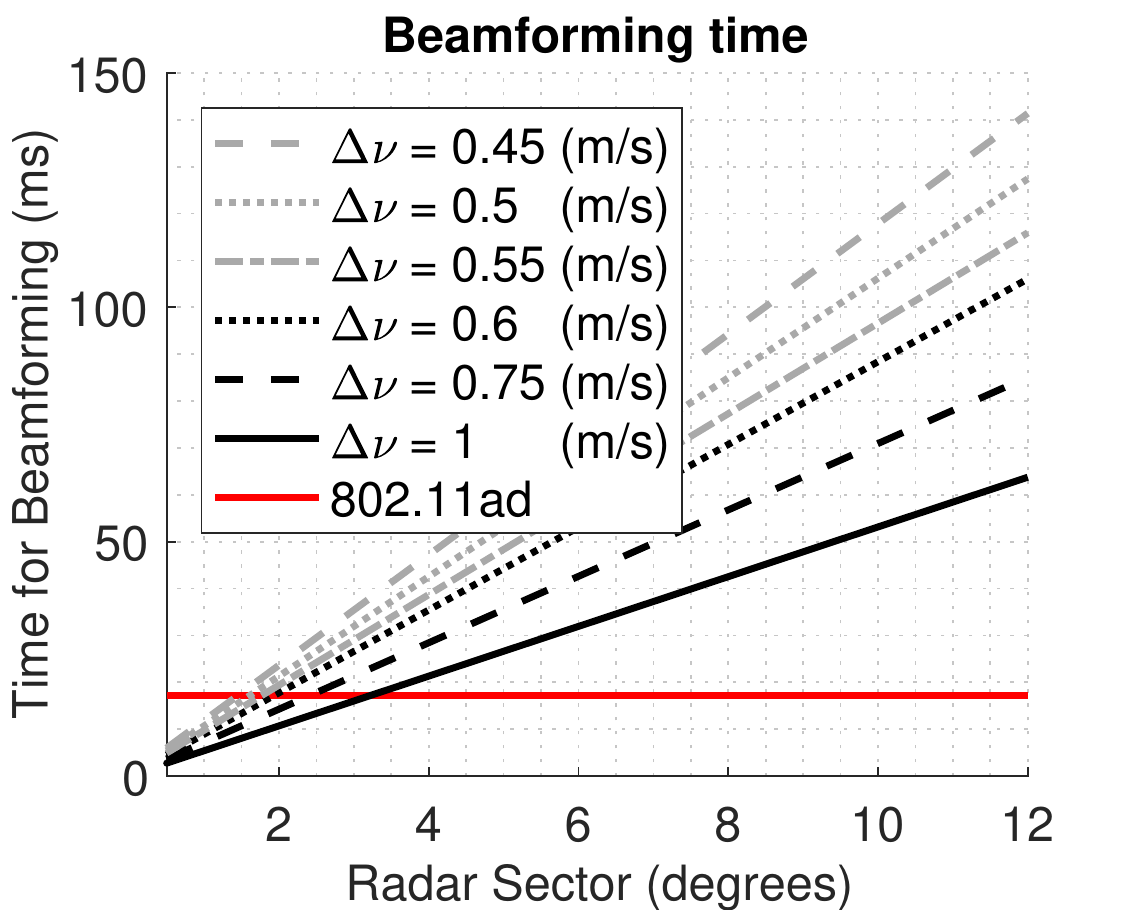}%
  	\label{subfig:OverheadvsSradar2}%
	}\qquad
	\subfloat[]{%
  	\includegraphics[width=0.45\linewidth]{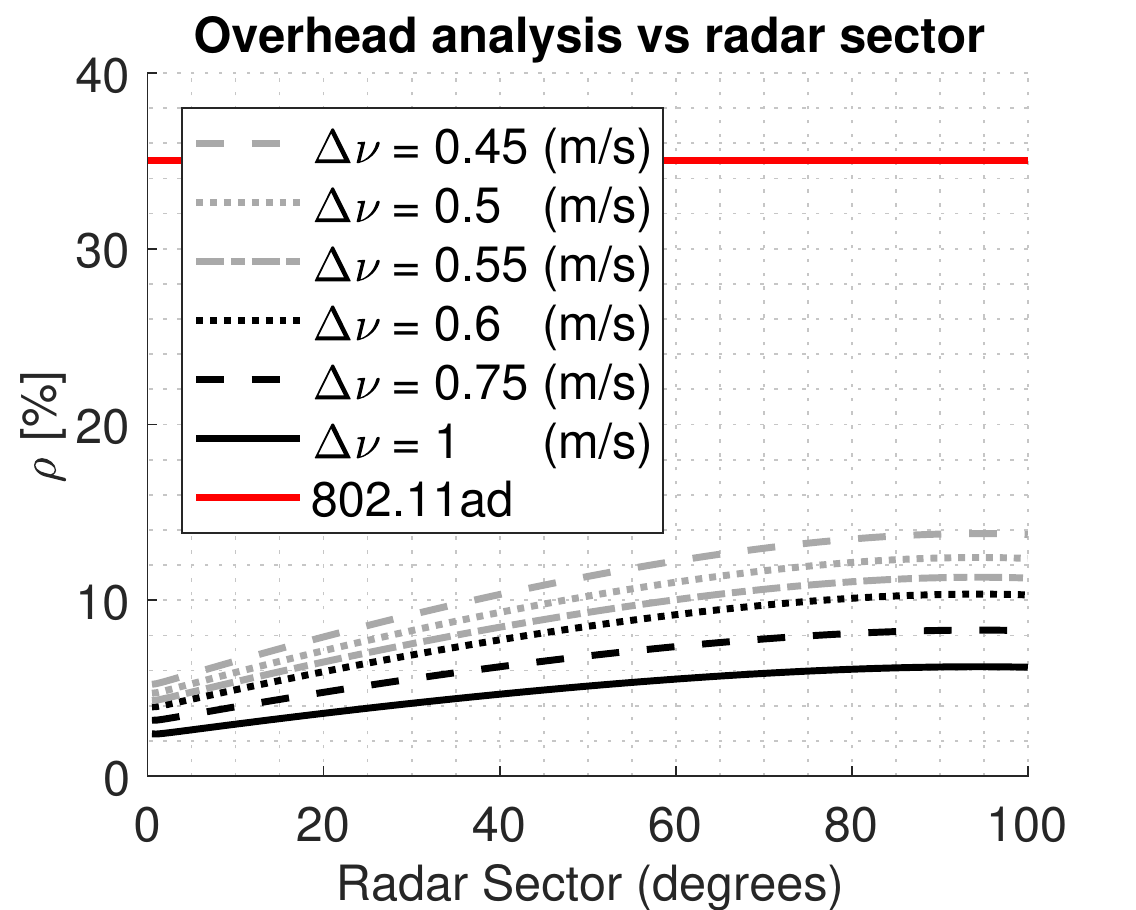}%
  	\label{subfig:rhovsSradar}%
	}
\vspace{1pt}
\caption{\small{(a) The time incurred for beamforming in one BHI worsens with radar. (b) The time spent performing beamforming is reduced because of its longer periodicity.}\vspace{-4pt}}
\label{fig:overhead}
\end{figure}

As stated in Section~\ref{sec:sys_desc}, we employ beam-overlapping to mitigate potential radar estimation errors. In Fig.~\ref{fig:CDFs}, the impact of the Doppler resolution ($\Delta \nu $) and beam-overlap on the misalignment probability is presented. Accurate velocity estimations require low $\Delta \nu$. However, as seen in Fig.~\ref{fig:dopres}, improved radar accuracy implies longer radar transmission periods.
We now study the beamforming time to observe the reduction in the overhead over baseline 802.11ad. Fig.~\ref{subfig:OverheadvsSradar2} shows the time a single beamforming procedure requires in terms of $\varphi_{\text{Sradar}}$. Even if the radar increases the beamforming time for a single BHI, as per (\ref{eq:T}), we only repeat this procedure every $T_{\text{radar}}$. In contrast, legacy 802.11ad for V2I requires such a repetition every $\approx$30 ms \cite{mavromatis2017beam}, which implies $\approx$35\% of the overall access time. Fig.~\ref{subfig:rhovsSradar} shows the percentage time (\ref{eq:rho2}) our configuration requires of the total time.

Consider a radar design with a $\Delta \nu$ of $0.45$ m/s and a $\varphi_{\text{Sradar}}$ of $5^{\circ}$. For a cumulative misalignment probability of less than 1\% at the end of the coverage area, an overlap ratio of 0.7 is employed. This implies a beam training overhead in the BHI of $59$ ms in contrast to the $10.72$ ms of the 802.11ad. However, our design introduces a total overhead of 5.82\% compared to the 35\% from the standard. Thus, we achieve a reduction of 83\% on the 802.11ad beam training overhead.

\vspace{-4pt}
\section{Conclusion}
\label{sec:concl}
This work demonstrates the feasibility for using radar and communications jointly within a single 802.11ad transceiver chain for high bandwidth and mobility situations. We configured the 802.11ad MAC to embed radar operations with  standards-compliant packets and at the same time, improve upon the completion time and scalability of legacy beam training procedures. We have studied the ideal parameter settings for the radar operation, including the impact of varying the sweep overlap and the achievable Doppler resolution. 
\vspace{-4pt}
\section*{Acknowledgments}
Kumar Vijay Mishra acknowledges partial support via Andrew and Erna Finci Viterbi Fellowship and Lady Davis Fellowship.
\vspace{-8pt}
\bibliographystyle{IEEEtran}
\bibliography{main}

\end{document}